\documentstyle[12pt]{article}

\def\sqr#1#2{{\vcenter{\hrule height.#2pt
      \hbox{\vrule width.#2pt height#1pt \kern#1pt
        \vrule width.#2pt}
      \hrule height.#2pt}}}

\def\abstracts#1#2#3{{
        \centering{\begin{minipage}{4.62in}\baselineskip=13pt
        \small
        \centerline{\bf Abstract}
        \vspace*{0.2cm}                % W. Janke (July 1, 1992)
        \parindent=0pt #1\par
        \parindent=18pt #2\par
        \parindent=15pt #3
        \end{minipage} }\par}}
\renewcommand{\thefootnote}{\fnsymbol{footnote}}
\begin{document}
\vspace*{-3cm}
\hfill \parbox{4.2cm}{ FUB-HEP 9/93\\
               May 4, 1993 \\
               }\\
\vspace*{1.5cm}
\centerline{\LARGE \bf Multicanonical Multigrid Monte
                       Carlo\footnotemark}\\[0.2cm]
\footnotetext{\noindent Work supported in part by Deutsche
Forschungsgemeinschaft under grant Kl256.}
\addtocounter{footnote}{-1}
\renewcommand{\thefootnote}{\arabic{footnote}}
\vspace*{0.4cm}
\centerline{\large {\em Wolfhard Janke\/}$^{1}$
               and {\em Tilman Sauer\/}$^2$}\\[0.4cm]
\centerline{\large    $^1$ {\small Institut f\"ur Physik,
                      Johannes Gutenberg-Universit\"at Mainz}}
\centerline{    {\small Staudinger Weg 7, 6500 Mainz 1, Germany }}\\[0.15cm]
\centerline{\large    $^2$ {\small Institut f\"{u}r Theoretische Physik,
                      Freie Universit\"{a}t Berlin}}
\centerline{    {\small Arnimallee 14, 1000 Berlin 33, Germany}}\\[2.50cm]
\vspace*{2.3cm}
\abstracts{}{
To further improve the performance of Monte Carlo simulations of
first-order phase transitions we propose to combine the multicanonical
approach with multigrid techniques. We report tests of this
proposition for the $d$-dimensional $\Phi^4$ field theory in two
different situations. First, we study quantum tunneling for $d = 1$ in
the continuum limit, and second, we investigate first-order phase
transitions for $d = 2$ in the infinite volume limit.
Compared with standard multicanonical simulations we obtain
improvement factors of several resp. of about one order of magnitude.
}{}
\thispagestyle{empty}
\newpage
\pagenumbering{arabic}

  At first-order phase transitions standard Monte Carlo simulations
in the canonical ensemble exhibit a supercritical slowing down
\cite{book}. Here extremely large autocorrelation times are caused
by strongly suppressed transitions between coexisting phases which,
on finite periodic lattices, can only proceed via mixed phase
configurations containing two interfaces.
Since the probability of such configurations is suppressed by a factor
$\exp(-2\sigma L^{d-1})$, where $\sigma$ is the interface tension
and $L^{d-1}$ the cross-section of the system, the autocorrelation
times in the simulation grow exponentially with the size of the
system, $\tau \propto \exp(2\sigma L^{d-1}$).

  A way to overcome this problem, known as umbrella \cite{umbrella}
or multicanonical \cite{muca1} sampling, is to simulate an auxiliary
distribution in which the mixed phase configurations have the same
weight as the pure phases and canonical expectations are computed by
reweighting \cite{wj92a}. Several tests for various models \cite{applic}
have demonstrated that this method works well in practice and reduces
supercritical slowing down to a power-like behavior
with $\tau \propto V^\alpha = L^{d\alpha}$, where $\alpha \approx 1$.
While this is clearly an important step forward the remaining slowing
down problem is still severe.
In most cases it is even worse than for standard (e.g., Metropolis or
heat-bath) Monte Carlo simulations of critical phenomena \cite{critical}.

  For the latter applications several update algorithms have been
developed which greatly reduce or even completely eliminate the
critical slowing down problem \cite{review}. In addition to
overrelaxation and cluster methods, an important class of such
algorithms are multigrid techniques \cite{mgmc1,mgmc2}. Here the
general strategy is to perform collective updates on different length
scales by visiting various coarsened grids in a systematic, recursively
defined way, generally known as V- or W-cycle \cite{mg_books}.

  Because of their conceptual simplicity both the multicanonical
reweighting approach and the multigrid update techniques are quite
generally applicable. The purpose of this note is to show that the two
approaches can easily be combined and give a much better performance
than each component alone.
We report tests of this combination for the $\Phi^4$ lattice field theory
with negative mass term in two conceptually different situations. We
first consider the quantum mechanical tunneling problem in one dimension
and study the performance of the new algorithm in the continuum limit.
We then discuss field driven first-order phase transitions in the
two-dimensional case and investigate the behavior of the multicanonical
multigrid algorithm in the infinite volume limit.

  For Potts models an interesting different approach was proposed only
recently in Ref.\cite{kari}. Here the idea is to combine a multicanonical
demon algorithm with cluster update methods in a hybrid-like fashion.

  The basic idea of the multicanonical approach is to sample the mixed
phase configurations with the same statistical weight as the
configurations of the pure phases. At a field driven first-order phase
transition this can always be achieved by a suitably chosen reweighting
factor $w^{-1}(m) \equiv \exp(-f(m))$, where $m=\sum_i \Phi_i/V$
is the average field. In a temperature driven transition, $m$ has
simply to be replaced by the average energy. Starting from an initial
guess based on experience or on some analytical approximation, a few
iterations are usually sufficient to adjust this factor. Once it is
fixed, canonical expectation values $\langle {\cal O} \rangle_{\rm can}$
of any observable $\cal O$ can be computed from the basic reweighting
formula
\begin{equation}
\langle {\cal O} \rangle_{\rm can} = \frac{ \langle w{\cal O}
\rangle}{\langle w \rangle},
\label{eq:rew}
\end{equation}
where $\langle \dots \rangle$ without subscripts denote expectation
values in the multicanonical distribution. To update field values with
a Metropolis algorithm in the multicanonical approach, we consider as
usual local moves $\Phi_i \rightarrow \Phi_i + \Delta \Phi_i$ and
compute the energy difference $\Delta E$. The decision of whether such
moves are accepted or not, however, is now based on the value of
$\Delta E + f(m+\Delta \Phi_i/V)-f(m)$.

  The basic idea of multigrid techniques is to perform updates on
different length scales. Using the so-called linear interpolation scheme
this amounts, in the equivalent unigrid viewpoint, to proposing moves
for blocks of $1, 2^d, 4^d,\dots, V=L^d=2^{nd}$ adjacent variables in
conjunction, with the sequence of length scales $2^{k}, k=0,\dots,n$
chosen in a specific, recursively defined order. Particular successful
sequences are the so-called V-cycle with
$k=0,1,\dots,n-1,n,n-1,\dots,1,0$  and the W-cycle whose graphical
representation looks like the letter W (for n=3, e.g., this is
0, 1, 2, 3, 2, 3, 2, 1, 2, 3, 2, 3, 2, 1, 0, and for large $n$ the W
looks more and more like a ``fractal''). In a canonical simulation
the update at level $k$ thus consists in considering a common move
$\Delta \Phi$ for all $2^{kd}$ variables of one block,
$\Phi_i \longrightarrow \Phi_i + \Delta \Phi$, $i \in {\rm block}$.

  The modifications for a multicanonical multigrid simulation are quite
trivial. Since at level $k$ the proposed move would change
the average field by $2^{kd} \Delta \Phi/V$, the decision of acceptance
is now simply to be based on the value of
$\Delta E + f(m + 2^{kd}\Delta \Phi/V) - f(m)$, with $\Delta E$
computed as in the canonical case. For didactic reasons we have
emphasized here the conceptually simpler {\em uni\/}grid viewpoint.
It should be stressed that in the recursive {\em multi\/}grid
formulation, which can be implemented more efficiently (similar to the
fast-Fourier transformation FFT), the multicanonical modification is
precisely the same.

  We have tested the multicanonical multigrid algorithm for the scalar
$\Phi^4$ lattice field theory in $d=1$ and $d=2$ dimensions,
defined by the partition function
\begin{equation}
Z = \prod_i^{L^d} \left[ \int_{-\infty}^{\infty} d\Phi_i/A \right]
\exp \left( -\epsilon \sum_{i=1}^{L^d} \left(
\frac{1}{2\epsilon^2} (\vec{\nabla} \Phi_i)^2
                       - \frac{\mu^2}{2} \Phi_i^2 + g \Phi_i^4 \right)
\right),
\label{eq:zphi4}
\end{equation}
with $A=\sqrt{2\pi\epsilon}$ and $\mu^2,g > 0$. We always impose
periodic boundary conditions. For $d=1$ we keep $L\epsilon = \beta$
fixed. Here the model describes the quantum statistics of a particle
tunneling back and forth in a double-well potential in contact with a
heat-bath at temperature $T = 1/\beta$ \cite{cf81}.
At fixed $\beta$ the limit $L \rightarrow \infty$
corresponds to the {\em continuum\/} limit. For $d \ge 2$ we put
$\epsilon = 1$. Here reflection symmetry is spontaneously
broken for all $\mu^2 > \mu_c^2(g) > 0$ as $L \rightarrow \infty$, which
is now the {\em infinite volume\/} limit. Consequently, if a term
$h\sum_i\Phi_i$ is added to the energy, the system exhibits a line of
first-order phase transitions driven by the field $h$.

Even though in the one-dimensional case no spontaneous symmetry breaking
occurs, the numerical difficulties are quite similar. This is due to
the fact that for small quartic coupling $g$ tunneling events are
strongly suppressed by a factor $\sim \exp(-const/g)$, which plays a
similar role as the factor $\exp(-2\sigma L^{d-1})$ at a first-order
phase transition. The important difference is, of course, that the
suppression factor stays roughly constant
in the continuum limit, while at a first-order phase transition it
rapidly decreases in the infinite volume limit. Nevertheless,
for small values of $g$ analogous slowing down problems in canonical
simulations of the quantum problem are notorious, and a number of
modified Monte Carlo schemes have been proposed in the past
\cite{negele,shuryak}. None of these techniques,
however, is general enough to be easily adapted to different potential
shapes.

  To evaluate the performance of the multicanonical multigrid algorithm,
we have recorded the time series for several observables and studied
their autocorrelation times. In this note we shall concentrate on the
average field $m = \sum_i \Phi_i/V$, which reflects most directly the
tunneling process.

  In previous investigations \cite{applic} emphasis was laid on the
{\em exponential\/} autocorrelation time $\tau^{(0)}_m$
of $m$, i.e., directly on the multicanonical dynamics. While this
nicely illustrated the absence of exponential slowing down, it is not
immediately clear how the remaining autocorrelations enter into the
error estimates for {\em canonical\/} expectation values computed
according to (\ref{eq:rew}).
To be precise, we are interested in the variance
$\epsilon^2 = \sigma^2_{\hat{\cal O}} = \langle \hat{\cal O}^2 \rangle
-\langle \hat{\cal O} \rangle^2$ of the (weakly biased) estimator for
$\langle \cal O \rangle_{\rm can}$, $\hat{\cal O} = \sum_1^{N_m} w(m_i)
{\cal O}_i/\sum_1^{N_m} w(m_i) \equiv \overline{w_i
{\cal O}_i}/\overline{w_i}$, if $N_m$ (multicanonical) measurements are
performed.
To facilitate a direct comparison with canonical simulations, we hence
{\em define\/} for multicanonical simulations an effective
autocorrelation time $\tau^{\rm eff}$ by the standard error formula for
$N_m$ correlated measurements,
\begin{equation}
\epsilon^2  = \sigma_{\rm can}^2 2\tau^{\rm eff}/N_m,
\label{eq:eff}
\end{equation}
where $\sigma_{\rm can}^2 = \langle {\cal O}_i^2 \rangle_{\rm can} -
\langle {\cal O}_i \rangle^2_{\rm can}$ is the variance of the canonical
distribution of single measurements, which can be computed in a
multicanonical simulation by using eq.~(\ref{eq:rew}). The squared error
$\epsilon^2$ can be estimated either by blocking or better by jack-knife
blocking \cite{jackknife} procedures, or by applying standard error
propagation to the variance of
$\hat{\cal O} = \overline{w_i {\cal O}_i}/\overline{w_i}$, which
involves the (multicanonical) variances and covariances of
$w_i {\cal O}_i$ and $w_i$, and the three associated autocorrelation
times $\tau_{m;m} \equiv \tau_m$, $\tau_{wm;wm} \equiv \tau_{wm}$, and
$\tau_{wm;m} = \tau_{m;wm}$ \cite{inprep}. By symmetry, for
${\cal O} = m$ this simplifies to
\begin{equation}
\epsilon^2 = \frac{\langle w_i m_i; w_i m_i \rangle}
                  {\langle w_i \rangle^2}
             \frac {2 \tau_{wm}}{N_m}
 \equiv \sigma^2_{\rm muca} \frac {2 \tau_{wm}}{N_m},
\end{equation}
where $\langle x;y \rangle \equiv \langle x y \rangle - \langle x \rangle
\langle y \rangle$ and $\tau_{x;y}=1/2 + \sum_k \langle x_0; y_k \rangle/
\langle x_0; y_0 \rangle$ is the integrated
autocorrelation time of multicanonical measurements.
In this way properties of the multicanonical distribution
(given by $\sigma^2_{\rm muca}$) are disentangled from properties of
the update algorithm (given by $\tau_{wm}$). Note that in
$\tau^{\rm eff} = (\sigma^2_{\rm muca}/\sigma^2_{\rm can}) \tau_{wm}$,
it is the autocorrelation time of $w(m) m$ that enters and not that of
$m$, as previously investigated.

  Let us first discuss our results for the quantum mechanical case in
$d=1$, where we shall confine ourselves to the case $\mu^2=1.0$,
$g=0.04$ and $\beta=10$. This choice of parameters may perhaps be
better characterized by the first few energy eigenvalues (obtained by a
numerical integration of the associated Schr\"odinger equation),
$E_0 = -0.913371$, $E_1 = -0.892348$, $E_2 = 0.029846$, and
$E_3 = 0.37813$, or the probability ratio
$P_{\rm min}/P_{\rm max} =P(0)/P_{\rm max} \approx 1.9 \times 10^{-3}$,
where $P(m)$ is the probability distribution of the magnetization (or,
in the present interpretation, of the average path). In a canonical
simulation it would consequently be about 500 times harder to sample
configurations with $m=0$ than configurations contributing to the peaks
of $P(m)$. To set up the multicanonical reweighting factor we started
from a variational approximation \cite{fk} that works quite well for
not too large $\beta$-values and is known to provide locally a lower
bound on $P(m)$ \cite{wj}.
Alternatively, as the distribution depends only weakly on $L$, one
could also use the distribution for small $L$ (which is quite
easy to generate by standard techniques) as input for the other
simulations.
Since the $m$-values vary continuously, we introduced bins of size
$\Delta m = 0.02$ to store the weight factor $w(m)$. A single short
run was usually sufficient to improve the initial guess such that the
multicanonical distribution $P'(m)$ had the desired flat shape,
$P'(m) \approx const$, between the two peaks.

In this way we performed multicanonical Metropolis and multigrid
simulations for $L=4,8,16,32,64,128$, and $256$, and using the
multigrid update also for $L=512$. In the multigrid case we
investigated both the V-cycle (using $n_{\rm pre}=n_{\rm post}=1$
pre- resp. post-sweeps \cite{mgmc1,mgmc2,mg_books}) and the W-cycle
(using $n_{\rm pre}=n_{\rm post}=1$ as well as $n_{\rm pre}=1$,
$n_{\rm post} = 0$). In the log-log plot of Fig.~1 we show
our results for $\tau^{\rm eff}$ of the multicanonical Metropolis and
W-cycle (without post-sweeps) update algorithm,
and for comparison also previous results \cite{js_mgmc} for the
canonical counterparts. We see that canonical and multicanonical
simulations exhibit qualitatively the same behavior in the continuum
limit $L \rightarrow \infty$. For both distributions, the Metropolis
update leads to a power-law growth $\tau \propto L^z$ with $z \approx 2$,
while for the W-cycle update the autocorrelation times stay roughly
constant \cite{tobepub}. Here it is the overall scale which is
reduced in the multicanonical simulation. For our choice of parameters
we obtain an improvement factor of about 60 (30) for the
Metropolis (W-cycle) update. For smaller $g$, since this factor is
essentially given by the inverse of the suppression factor
$\exp(-const/g)$, we found the multicanonical Metropolis update to be
more favorable than the canonical W-cycle for reasonably large $L$.
Eventually, however, there will always be a crossover at some $L$.
For $g = 0.04$, if we compare the Metropolis update and the W-cycle
in the multicanonical simulation we obtain an improvement
factor of about $50,000$ for $L=512$.

  In the continuum limit the distribution $P(m)$ and the ratio
$P_{\rm min}/P_{\rm max}$ stay roughly constant. This implies that the
multicanonical approach can only give an improvement factor which is
independent of $L$. In this case it is the update algorithm that plays
the dominant role asymptotically for large $L$, while the multicanonical
reweighting procedure sets the overall scale. The relative importance
of the two approaches is just reversed at first-order phase
transitions when the infinite volume limit is considered.

  As a test case we have studied the model (\ref{eq:zphi4}) with
$\epsilon=1$. For this model the line of second-order phase transitions
separating the broken and unbroken phase in the $\mu^2-g$ plane has
recently been determined by Toral and Chakrabarti \cite{toral}.
Here we concentrate on the first-order phase transition between the two
ordered phases at $g=0.25$ and $\mu^2=1.30$, which is sufficiently far
away from the critical point at $\mu^2_c = 1.265(5)$ \cite{toral} to
display the typical behavior already on quite small lattices.
A sensitive measure of the strength of the transition is the interface
tension $\sigma_{oo}$ between the $+$ and $-$ phase, which turns
out \cite{inprep} to be $\sigma_{oo} = 0.03459(49)$
(for comparison, about the same value is found
for the order-disorder interface tension in the two-dimensional $q$-state
Potts model with $q=9$, where $\sigma_{od}=0.03355\dots$ \cite{bj92}).
We performed multicanonical simulations using the Metropolis update and
the W-cycle without post-sweeps for lattices of size
$V=L^2$ with $L=8,16$ and $32$. With the multigrid algorithm, due to
the improved performance, we were also able to study lattices of
size $L=64$.
Here each time series contains a total of $10^6$ measurements taken
every $n_e${\em th} sweep, after discarding $10^4 \times n_e$ sweeps
for thermalization. The number of sweeps between measurements, $n_e$,
was adjusted in such a way that in each simulation the measurements of
$wm$ had an autocorrelation time of maximal 50, i.e., the length of each
time series is at least $20,000~\tau_{wm}$.

  A few of our results are collected in Table~1, where we give the
integrated and exponential autocorrelation times of $m$ and $w(m)m$ as
well as $\tau^{\rm eff}$ according to eq.~(\ref{eq:eff}) for both update
algorithms. We see that integrated and exponential autocorrelation
times for $m$ agree well with each other, showing that the corresponding
autocorrelation function can be approximated by a single exponential. For
$wm$ we obtain values for $\tau^{(0)}$ that are consistent with those for
$m$ within error bars. The integrated autocorrelation times, however, are
significantly lower, implying that the autocorrelation function is
composed of many different modes. We also observe that the
difference between $\tau_{wm}$ and $\tau^{\rm eff}$ can be quite
appreciable. From $L=8$ to $L=64$ the ratio
$\tau^{\rm eff}/\tau_{wm} = \sigma^2_{\rm muca}/\sigma^2_{\rm can}$
varies from about $1.9$ to $4.6$, reflecting the varying probability
distribution shapes with increasing $L$.

  By fitting $\tau^{\rm eff}$ to a power law,
$\tau^{\rm eff} \propto L^z$, we obtain for both update algorithms an
exponent of $z \approx 2.3$, i.e., in this case it is thus the multigrid
update that reduces the overall scale. The autocorrelation times of the
W-cycle are reduced by a roughly constant factor of about 20 as compared
with the Metropolis algorithm. Of course, for a fair comparison we should
also take into account that a W-cycle requires more elementary
operations than a Metropolis sweep  \cite{mgmc1}.
Such a work estimate, however, depends on many details of the
implementation and it is hence difficult to give generally valid
figures.
With our implementations on a CRAY Y-MP we obtained a {\em real time}
improvement factor of about 10.
\begin{table}[hb]
 \begin{center}
 \caption[a]{Autocorrelation times for the multicanonical simulation
using the standard Metropolis (M) or multigrid W-cycle (W) update
algorithm.\\}
% -----------------------------------------------------
% adapted from TeX book, p. 241
\newlength{\digitwidth} \settowidth{\digitwidth}{\rm 0}
\catcode`?=\active \def?{\kern\digitwidth}
% -----------------------------------------------------
  \begin{tabular}{|r|r|l|l|l|l|}
\hline
\multicolumn{1}{|r|}{ }      &
\multicolumn{1}{c|}{ }       &
\multicolumn{1}{c|}{$L=8$}  &
\multicolumn{1}{c|}{$L=16$} &
\multicolumn{1}{c|}{$L=32$} &
\multicolumn{1}{c|}{$L=64$}\\
\hline
$\tau^{(0)}_{m}$  & M & 212(12)    & ?668(23)    & 3120(200)   & $-$         \\
                    & W & ?11.30(32) & ??37.2(2.0) & ?148(11)    &  ?746(62) \\
$\tau_{m}$        & M & 204.4(4.0) & ?690(11)    & 2984(63)    & $-$         \\
                    & W & ?10.88(12) & ??34.69(76) & ?150.0(4.0) &  ?758(37) \\
\hline
$\tau^{(0)}_{wm}$& M & 209(12)    & ?655(31)    & 2880(190)   & $-$         \\
                    & W & ?11.34(33) & ??36.9(2.0) & ?146(13)    &  ?600(120)\\
$\tau_{wm}$      & M & 171.1(3.4) & ?509.8(8.9) & 1840(40)    & $-$         \\
                    & W & ??9.82(11) & ??27.58(59) & ??96.6(2.4) &  ?374(23) \\
\hline
$\tau^{\rm eff}$    & M & 322.7(6.1) & 1258(21)    & 6050(120)   & $-$
\\
                    & W & ?18.51(20) & ??67.4(1.3) & ?321.9(7.6) & 1724(86)  \\
\hline
   \end{tabular}
  \end{center}
\end{table}
%
%----------------------------------------------------------------
       \section*{Acknowledgments}
%----------------------------------------------------------------
%
W.J. thanks the Deutsche Forschungsgemeinschaft for a Heisenberg
fellowship.
%
%-----------------------------------------------------------------------
               
%
\newpage
%
%----------------------------------------------------------------
  {\Large\bf Figure Heading}
%----------------------------------------------------------------
%
  \vspace{1in}
  \begin{description}

    \item[\tt\bf Fig. 1:]
Effective autocorrelation times $\tau^{{\rm eff}}$ for the model (2)
in $d = 1$ as a function of lattice size $L$ with
$L\epsilon = \beta = 10$, $\mu^2 = 1$, $g = 0.04$
for different Monte Carlo algorithms. The canonical data are taken from
Ref. \cite{js_mgmc}.

 \end{description}

\begin{thebibliography}{19}
%-----------------------------------------------------------------------
%
\bibitem{book}
For recent reviews, see, e.g.,
{\em Dynamics of First Order Phase Transitions},
edited by H.J. Herrmann, W. Janke, and F. Karsch
%proceedings of the
%workshop, J\"ulich, 1-3 June 1992, edited by
(World Scientific, Singapore, 1992).
%
\bibitem{umbrella}
G.M. Torrie and J.P. Valleau, {\em Chem. Phys. Lett.} {\bf 28} (1974)
578; {\em J. Comp. Phys.} {\bf 23} (1977) 187;
I.S. Graham and J.P. Valleau, {\em J. Phys. Chem.} {\bf 94} (1990) 7894;
J.P. Valleau, {\em J. Comp. Phys.} {\bf 96} (1991) 193.
%
\bibitem{muca1}
B.A. Berg and T. Neuhaus, {\em Phys. Lett.} {\bf B267} (1991) 249.
For a review, see B.A. Berg, in Ref.\cite{book}, p.311 [reprinted in
{\em Int. J. Mod. Phys.} {\bf C3} (1992) 1083].
%
\bibitem{wj92a}
W. Janke in Ref.\cite{book}, p.365 [reprinted in
{\em Int. J. Mod. Phys.} {\bf C3} (1992) 1137]; and preprint HLRZ 57/92,
J\"ulich (1992).
%
\bibitem{applic}
B.A. Berg and T. Neuhaus, {\em Phys. Rev. Lett.} {\bf 68} (1992) 9;
W. Janke, B.A. Berg, and M. Katoot,
{\em Nucl. Phys.} {\bf B382} (1992) 649;
B.A. Berg, U. Hansmann, and T. Neuhaus,
{\em Phys. Rev.} {\bf B47} (1993) 497;
{\em Z. Phys.} {\bf B90} (1993) 229;
A. Billoire, B.A. Berg, and T. Neuhaus, preprint SPhT-92/120,
Saclay (1992);
B. Grossmann and M.L. Laursen, in Ref.\cite{book}, p.375 [reprinted in
{\em Int. J. Mod. Phys.} {\bf C3} (1992) 1147];
and preprint HLRZ 7/93, J\"ulich (1993);
B. Grossmann, M.L. Laursen, T. Trappenberg, and U.-J. Wiese,
{\em Phys. Lett.} {\bf B293} (1992) 175.
%
\bibitem{critical}
K. Binder, in {\em Monte Carlo Methods in Statistical Physics\/},
edited by K. Binder (Springer, New York, 1979), p.1.\\
See also the articles in
{\em Finite-Size Scaling and Numerical Simulations of Statistical
Systems\/}, edited by V. Privman (World Scientific, Singapore, 1990).
%
\bibitem{review}
%
R.H. Swendsen, J.-S. Wang, and A.M. Ferrenberg, in
{\em The Monte Carlo Method in Condensed Matter Physics\/}, edited
by K. Binder (Springer, Berlin, 1991);
%
C.F. Baillie, {\em Int. J. Mod. Phys.} {\bf C1} (1990) 91;
A.D. Sokal, {\em Monte Carlo Methods in Statistical
Mechanics: Foundations and New Algorithms\/}, Cours de
Troisi\`eme Cycle de la Physique en Suisse Romande,
Lausanne, 1989.
%
\bibitem{mgmc1}
%
J. Goodman and A. D. Sokal, {\em Phys. Rev. Lett.} {\bf 56}, 1015 (1986);
                         {\em Phys. Rev.} {\bf D40}, 2035 (1989).
%
\bibitem{mgmc2}
%
D. Kandel, E. Domany, D. Ron, A. Brandt, and E. Loh, Jr.,
        {\em Phys. Rev. Lett.} {\bf 60}, 1591 (1988);
D. Kandel, E. Domany, and A. Brandt,
        {\em Phys. Rev.} {\bf B40}, 330 (1989).
%
\bibitem{mg_books}
%
W. Hackbusch, {\em Multi-Grid Methods and Applications\/}
(Springer, Berlin, 1985);
S.F. McCormick (ed.),
{\em Multigrid Methods. Theory, Applications, and Supercomputing\/}
(Dekker, New York, 1988).
%
\bibitem{kari}
K. Rummukainen, {\em Nucl. Phys.} {\bf B390} (1993) 621.
%was  preprint CERN-TH6654/92
%
\bibitem{cf81}
M. Creutz and B. Freedman, {\em Ann. Phys.} {\bf 132} (1981) 427.
For reviews see,
e.g., B.J. Berne and D. Thirumalai, {\em Ann. Rev. Phys. Chem.} {\bf 37}
(1986) 401;
N. Makri, {\em Comp. Phys. Comm.} {\bf 63} (1991) 389.
%
\bibitem{negele}
C. Alexandrou, Ph.D. Thesis, MIT, Cambridge (1985);
C. Alexandrou and J.W. Negele, {\em Phys. Rev.} {\bf C37} (1988) 1513.
%
\bibitem{shuryak}
E.V. Shuryak and O.V. Zhirov,
   {\em Nucl. Phys.} {\bf B242} (1984) 393;
E.V. Shuryak,
{\em Usp. Fiz. Nauk.} {\bf 143} (1984) 309
[{\em Sov. Phys. Usp.} {\bf 27} (1984) 448].
%
\bibitem{jackknife}
R.G. Miller, {\em Biometrika} {\bf 61} (1974) 1; B. Efron,
{\em The Jackknife, the Bootstrap and other Resampling Plans\/}
(SIAM, Philadelphia, PA, 1982).
%
\bibitem{inprep}
W. Janke and T. Sauer, in preparation.
%
\bibitem{fk}
R. Giachetti and V. Tognetti, {\em Phys. Rev. Lett.} {\bf 55} (1985) 912;
R.P. Feynman and H. Kleinert, {\em Phys. Rev.} {\bf A34} (1986) 5080;
W. Janke and H. Kleinert, {\em Chem. Phys. Lett.} {\bf 137} (1987) 162.
For a comprehensive review, see H. Kleinert,
{\em Path Integrals in Quantum Mechanics,
Statistics and Polymer Physics} (World Scientific, Singapore, 1990).
%
\bibitem{wj}
W. Janke, in {\em Path Integrals from meV to MeV\/}, proceedings,
Bangkok, 1989, ed. V. Sa-yakanit {\em et al.} (World Scientific,
Singapore, 1989).
%
\bibitem{js_mgmc}
%
W. Janke and T. Sauer, {\em Chem. Phys. Lett.} {\bf 201} (1993) 499;
and preprint HLRZ 108/92, to be published in {\em Path Integrals from
meV to MeV\/}, proceedings, Tutzing, 1992.
%
\bibitem{tobepub}
%
Using the V-cycle the exponent $z$ can only be reduced to
$z \approx 1$ for both distributions.
Also, the extra work required for the W-cycle with
$n_{pre} = n_{post} = 1$ is not completely
balanced by reduced autocorrelations.
W. Janke and T. Sauer, to be published.
%
\bibitem{toral}
%
R. Toral and A. Chakrabarti, {\em Phys. Rev.} {\bf B42} (1990) 2445;
see also
A. Milchev, D.W. Heermann, and K. Binder, {\em J. Stat. Phys.} {\bf 44}
(1986) 749.
%
\bibitem{bj92}
%
C. Borgs and W. Janke, {\em J. Phys. I (France)} {\bf 2} (1992) 2011.
%
\end{thebibliography}
\end{document}